\def\mag{\hbox{$\;.\!\!\!^m$}}
\def\muspc{\hskip 0.15 em}
\def\fdg{\hbox{$.\!\!^\circ$}}
\input psfig
\documentstyle[proceedings]{crckapb}

\begin{opening}
\title{Stellar population synthesis}
\subtitle{Baade's Window: OGLE vs DeNIS}

\author{Y.K. Ng}
\institute{Institut d'Astrophysique de Paris -- CNRS\\
           98bis, Boulevard Arago, F-75014 \ Paris, FRANCE}
\end{opening}

\begin{document}

\begin{abstract}
Monte-Carlo simulations of optical and near-infrared 
colour-magnitude diagrams (CMDs) in Baade's Window 
(\hbox{$l\!=\!1\fdg0,b\!=\!-3\fdg9$})
are compared with each other. 
The morphological structure of the red horizontal branch
in those CMDs is likely due to a combination of extinction
and age-metallicity of the stars from the bulge/bar.
In contrast with the optical (V,I) passbands,
the simulations indicate that it is feasible
to separate metallicity from extinction with
near-infrared data.
\end{abstract}

\section{Introduction}
Synthetic Hertzsprung-Russell Diagrams (HRDs) are generated through
the population synthesis technique from a homogeneous
library of stellar evolutionary tracks (Bertelli et al. 1994).
Figure 1 shows a schematic diagram of the HRD-GST 
(see Ng 1994 and Ng et al. 1995 for details) which uses these
diagrams as input for star counts studies.
With the HRD-GST the contribution of various stellar populations
are decomposed statistically.
Through a detailed analysis of the star counts along the line 
of sight the aim of the study is to obtain constraints on
\begin{list}{$\bullet$}{%
\topsep=0pt
\itemsep=0pt 
\parsep=0pt}
\item{the galactic structure;}
\item{the interstellar extinction; and}
\item{the age \& metallicity of the different stellar populations.}
\end{list}
The results obtained thus far have been
reported in the Papers I--III (Ng et al.
1995 \& 1996a and Bertelli et al. 1995)
and Bertelli et al. (1996). \hfill\break
The distribution of stars along the line of sight is the 
result of a complex mixture of populations.
The ages, metallicities and the spatial distributions of the 
stars from the various populations contain a wealth of 
information about the formation and evolution of our Galaxy.
\par

\setbox1=\vbox{\hsize=5.6cm
\null\noindent\quad
\psfig{file=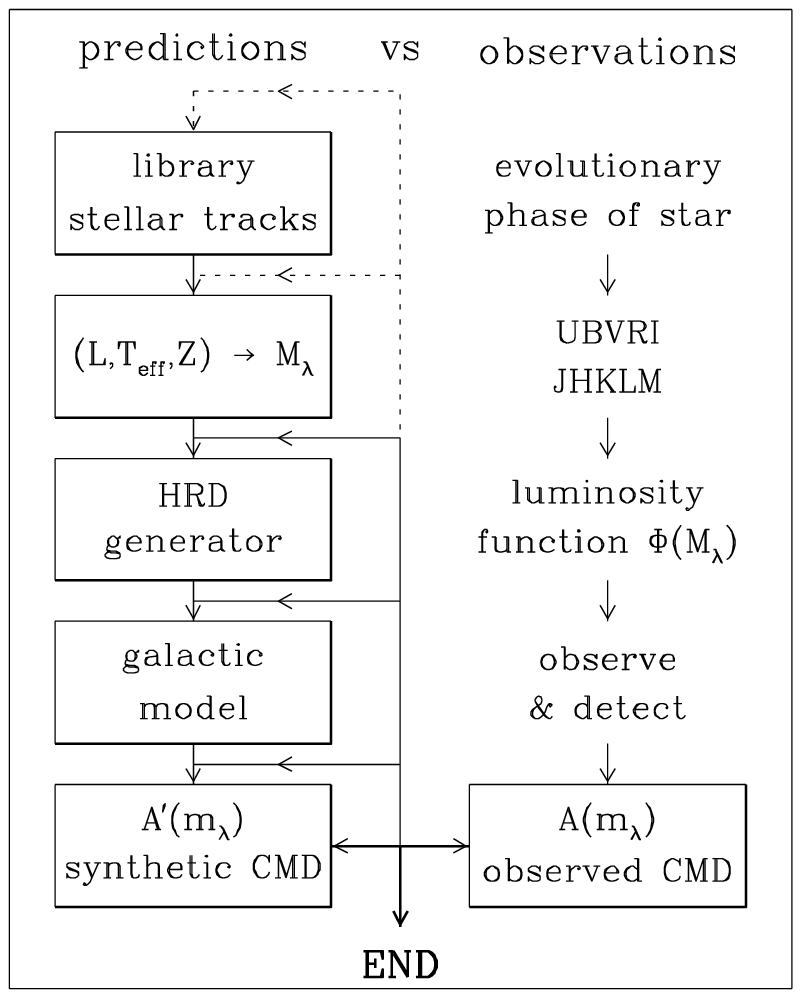,height=5.9cm,width=5.5cm}
}

\setbox2=\vbox{\hsize=5.7cm
\footnotesize
\noindent {\bf Figure 1.}\quad 
Schematic diagram of the HRD-GST. Input for the 
stellar population synthesis engine is the Padova library of
stellar evolutionary tracks 
(see Paper I or Bertelli et al. 1994 and references cited in those 
papers for details). 
The luminosities and effective temperature for each
synthetic star of arbitrary metallicity 
is then transformed to an absolute
magnitude in a photometric passband with the method outlined by
Bressan et al. (1994).
A synthetic HR-diagram is generated,
after specification of the stellar luminosity function 
through 
the initial mass function, the star formation rate and the
age \& metallicity range.
Synthetic stars from those diagrams are then `observed'
and `detected' with the galac-}

\vbox{%
\centerline{\hfil\copy1\quad\quad\quad\copy2}
\footnotesize
\noindent
tic model, through a Monte-Carlo technique.
In this model the density distribution of each galactic component along the 
line of sight is specified.
This results in a synthetic CMD of the field of interest.
The synthetic CMD ought to be 
comparable with the observed CMD,
when a realistic set of input parameters is used.
If there is a marginal agreement
then check the input for each step of the HRD-GST. 
%In an optimistic case the difference is due to 
%normalization. In the worst case the input library of 
%stellar evolutionary tracks is not sufficient.
}

\section{Extinction vs metallicity}
Towards the galactic centre there is no consensus
about the ages \& metallicities of the stellar
populations (disc, bulge, bar ...) and the parameterization of 
the galactic structure.
The large variation of the extinction over a relative small area is 
one of the major causes. Mainly, because it is not easily parameterized.
Therefore the CMDs of each field/region needs to be studied
separately for its extinction along the line of sight.
\par
In (V,V--I) CMDs the effects of extinction and metallicity 
are difficult to separate from each other.
Because the extinction vector points almost
in the same direction as the metallicity gradient, if
present, in the HB stars (see figure 1).
The tilted clump
of red HB stars could be caused by large differential extinction,
a large metallicity spread of the HB stars or a combination
of both 
(Catalan \& de Freitas Pacheco 1996; Ortolani et al. 1990 \& 1995;
Ng et al. 1996a,b\muspc\&\muspc{c}).
\par
This might imply that there are no super-metal rich
stars present towards the galactic centre.
Patchy extinction might explain the various structures present 
in the (V,V--I) CMDs. A study of the effects of various
forms of extinction towards Baade's Window (Ng \& Bertelli 1996)
shows that, if the extinction is not extremely high,
patchy extinction will not give results 
significantly different from a Poissonian type of extinction.
The morphology of the red HB stars in Baade's 
Window cannot be explained solely by (patchy) extinction
and a contribution due to metallicity ought to be taken into account.

\setbox3=\vbox{\hsize=5.6cm
\null\noindent\quad
\psfig{file=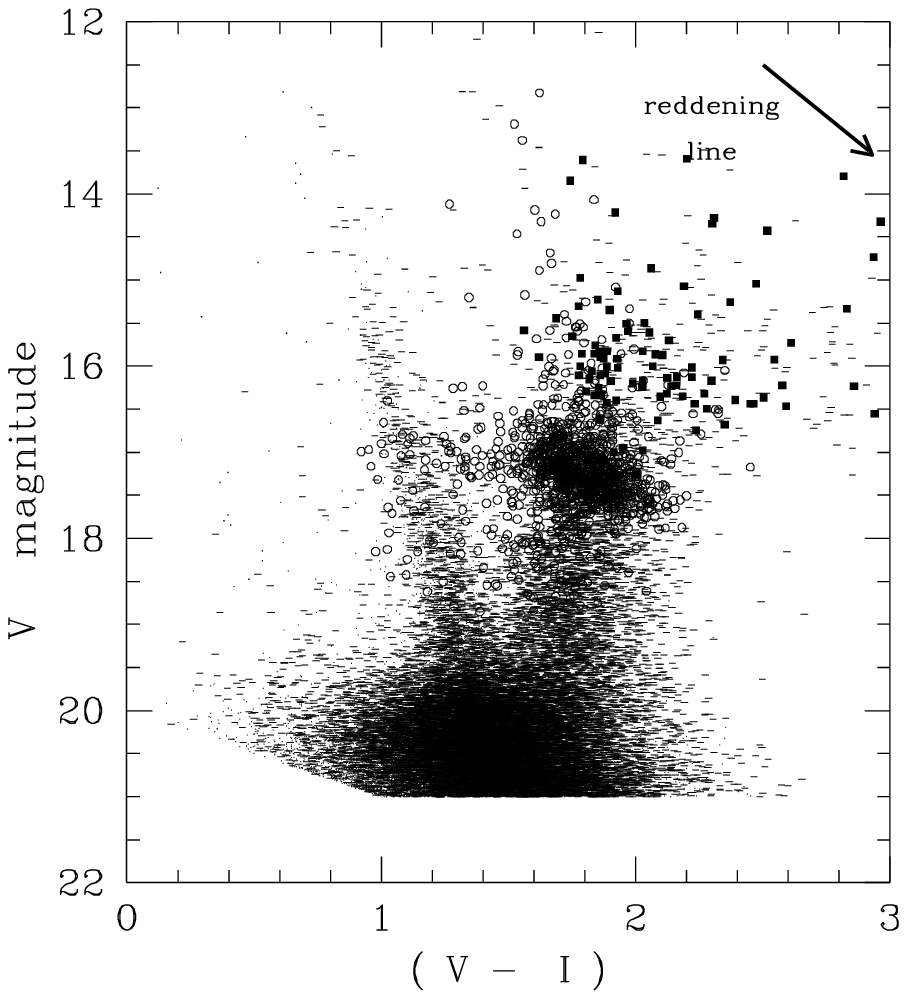,height=5.9cm,width=5.5cm}
}

\setbox4=\vbox{\hsize=5.6cm
\null\noindent\quad
\psfig{file=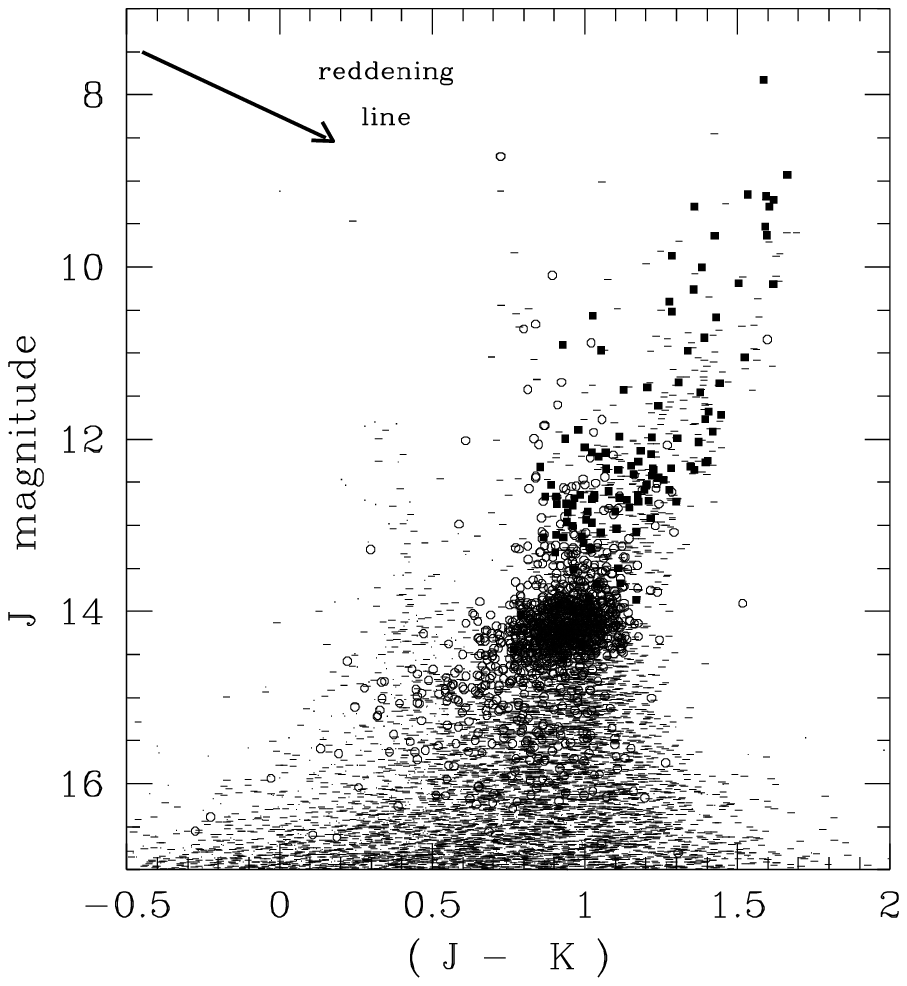,height=5.9cm,width=5.5cm}
}

\setbox5=\vbox{\hsize=5.3cm
\footnotesize
\noindent {\bf Figure 2.}\quad 
Simulated (V,V--I) CMD for Baade's Window
(subfield \#3)
\null\hfill\break
}

\setbox6=\vbox{\hsize=5.3cm
\footnotesize
\noindent {\bf Figure 3.}\quad 
Simulated (J,J--K) CMD for Baade's Window
(subfield \#3)
\null\hfill\break
}
\bigskip
\vbox{%
\centerline{\copy3\quad\copy4}
\vskip0.5cm
\centerline{\quad\quad\copy5\quad\quad\copy6\quad}
}

\section{Optical vs near-IR}
Although unlikely, patchy extinction with very special conditions
cannot be fully excluded.
On the other hand,
the near-IR passbands are less sensitive 
to extinction. The best strategy is probably to determine
the extinction (if this is not too high and/or patchy)
\& the metallicity range of the stellar populations 
from the optical passbands and verify the results with 
near-IR photometry.
\par
Baade's Window is used as an example, because 
this field has been studied in detail by the OGLE
collaboration (Udalski et al. 1993 and references 
cited therein). 
Furthermore detailed studies have been made of 
its CMD (Paczy\'nski et al. 1994, Ng et al. 1996a)
and of 
its extinction (Ng \& Bertelli 1996a, Stanek 1996).
From the study of the \hbox{(V,V--I)}
CMD Ng et al. (1996a\&b) found indications
for the age-metallicity of the `bar' population 
along the line of sight. It is not clear though 
how much the results in those studies are affected by 
under- or overestimating either the effect of extinction
over the field or the metallicity range of the stars
from various stellar populations.
\par
Figure 2 shows the simulated (V,V--I) CMD for this
window with the normalization reported in Paper III.
Note however, that the blue HB stars 
(\hbox{V\muspc$\simeq$\muspc17$^m$}) are slightly too faint 
with respect to the observed CMD. This is likely
due to the lower limit of the metallicity (\hbox{Z\muspc=\muspc0.0004})
adopted in the current simulation. With an even lower value of the
metallicity the blue HB stars will be slightly brighter.
The blue edge of the red HB is due to stars with metallicity
\hbox{Z\muspc=\muspc0.003}, while the metallicity of the HB stars
at the red edge could be as high as \hbox{Z\muspc=\muspc0.06}.
The latter value might be considered as an upper limit for 
Baade's Window.
This figure demonstrates that the reddening
vector in the (V,V--I) CMD is indeed almost parallel to 
the direction in which the metallicity increases.
\par
Figure 3 shows the (J,J--K) CMD. The extinction is 
scaled from the visual with the ratios A$_\lambda$/A$_V$
given by Rieke \& Lebofsky (1985).
It demonstrates that in the near-IR the extinction
and metallicity vector are not coupled anymore.
The simulation indicates that a more detailed 
study of the extinction and the metallicity of 
the stars should be feasible with the DeNIS survey
(Epchtein et al. 1993), taking into account the limiting magnitudes 
\hbox{J$_{lim}$\muspc$\simeq$\muspc16\mag5} and 
\hbox{K$_{lim}$\muspc$\simeq$\muspc14\mag5}.
\par
With near-infrared data it is possible to disentangle 
extinction from metallicity. A well constrained metallicity
range would provide important clues about the star formation history 
of our Galaxy (Ng \& Bertelli 1996b). 
Furthermore, it might provide indications 
about the presence or absence of super-metal rich stars,
which could improve our understanding of the nature of
the UV excess in elliptical galaxies (Bressan et al. 1994, 
Bertelli et al. 1996)

\bigskip
\noindent
{{\it Acknowledgements}\quad
I am indebted to the Padova group for providing the
stellar population synthesis code.
Ng is supported by HCM grant
CHRX-CT94-0627 from the European Community.}

\end{document}